\documentclass{article}
 \usepackage{graphicx}

 \title{Lie point symmetries \\ of difference equations and lattices}
\author{D. Levi\thanks{Dipartimento di Fisica, Universit\`a Roma Tre and
INFN--Sezione
di Roma Tre, Via della Vasca Navale 84, 00146 Rome, Italy}
 \and S. Tremblay\thanks{Centre
de Recherches Math\'ematiques and D\'epartement
 de Physique, Universit\'e de Montr\'eal, C.P. 6128, succ. Centre-ville,
Montr\'eal (QC), H3C 3J7, Canada} \and P. Winternitz\thanks{Centre de Recherches
Math\'ematiques and D\'epartement de Math\'ematiques et Statistiques,
Universit\'e de Montr\'eal,
 C.P. 6128, succ. Centre-ville, Montr\'eal (QC), H3C 3J7, Canada}}

 \date{}

 \def\neq {\not\equiv}

 \def\be   {\begin{equation}}   \def\ee   {\end{equation}}
 \def\ba   {\begin{array}}      \def\ea   {\end{array}}
 \def\bea  {\begin{eqnarray}}   \def\eea  {\end{eqnarray}}
 \def\bean {\begin{eqnarray*}}  \def\eean {\end{eqnarray*}}

\newcommand{\me}{\mathrm{e}}

\begin{document}

\maketitle

\begin{abstract}
A method is presented for finding the Lie point symmetry transformations acting
simultaneously on difference equations and lattices, while leaving the solution
set of the
corresponding difference scheme invariant. The method is applied to several
examples. The
found symmetry groups are used to obtain particular solutions of
differential-difference
equations.
\end{abstract}

\section{Introduction}

Lie groups have long been used to study differential equations. As a matter of
fact,
they
originated in that context \cite{1,2}. They have been put to good use to solve
differential
equations, to classify them, and to establish properties of their solution spaces
\cite[\ldots, 8]{3}.

Applications of Lie group theory to discrete equations, like difference equations,
differential-difference equations, or $q$-difference equations are much more
recent
\cite[\ldots, 37]{9}.

Several different approaches are being pursued. One philosophy is to consider a
given
system of discrete equations on a given fixed lattice and to search for a group of
transformations, taking solutions into solutions, while leaving the lattice
invariant.
Within this philosophy different approaches differ by the restrictions imposed on
the
transformations and by the methods used to find the symmetries. One thing that is
clear
is that within this philosophy it is necessary to generalize the concept of point
symmetries for difference equations, if we wish to recover all point symmetries of
a
differential equation in the continuous limit
\cite[\ldots, 26]{9}.

A different philosophy is to consider a difference equation and a lattice as two
relations involving a fixed number of points, in which we give the values of the
independent and dependent variables say $x_{-},x,x_{+}$ and $u_{-},u,u_{+}$
respectively.
The group transformations act on the equation and on the lattice. This
philosophy
was mainly developped by Dorodnitsyn and collaborators
\cite[\ldots, 33]{27}. In this approach, the given object
was a Lie group and its Lie algebra. Invariants of this Lie group, depending on
$x$
and
$u$, calculated at a predetermined number of points were obtained. They were used
to
obtain invariant equations and lattices. The emphasis was on discretizing
differential
equations while preserving all of their point symmetries, or at least most of
them.

The purpose of this article is to combine the two philosophies. More specifically,
we
will consider given equations on given lattices, but the lattice will also be
given
by
some equation. We will then look for Lie point transformations, acting on both
equations,
and leaving the common solution sets of both equations invariant.

In Section 2  we develop the formalism necessary for calculating simultaneous
symmetries
of difference or differential-difference equations and lattices. Section~3 is
devoted
to
examples of symmetries of purely difference equations, both linear and nonlinear
ones. In
 Section~4 we also consider examples, this time of differential-difference
equations.
Some conclusions are drawn in the final Section~5.

\section{Symmetries of differential-difference equations}

\subsection{The differential-difference scheme}

In this article we shall only consider a restricted class of problems, for reasons
of
simplicity and clarity. However, the formalism involved can easily be extended to
quite
general systems of equations.

Thus we shall consider one scalar function $u(x,t)$ of two variables only. The
variable
$t$ is continuous and varies in some interval $I\subset \Re$. The variable
$x$
is
also continuous and varies in some interval $\tilde I\subset \Re$. However,
$x$
will be `sampled' in a set of discrete points
$\{\ldots,x_{n-2}\, ,\, x_{n-1}\, ,\, x_{n}\, ,\, x_{n+1}\, ,\ldots\}$. The points
$x_{k}$ are not
necessarily
equally spaced.

We shall study the symmetries of a pair of equations which we postulate to have
the
form
\begin{eqnarray}
E=E\left(t,\{x_{k}\}_{k=n-n_{1}}^{n+n_{2}},\{u_{k}\}_{k=n-n_{1}}^{n+n_{2}},u_{n,t}
,u_
{n,t
t}\right)&=&0 \label{eq:E=0}
\\*[2ex]
\Omega=\Omega\left(t,\{x_{k}\}_{k=n-n_{3}}^{n+n_{4}},\{u_{k}\}_{k=n-n_{3}}^{n+n_{4
}}\right)&=&0\ \ ,\ \ n_{i}\ge 0\,.
\label{eq:omega=0}
\end{eqnarray}

We have $k,n,n_{i}\in {Z}$, all $n_{i}$ are finite. Equations
(\ref{eq:E=0})
is a
differential equation in $t$ and a difference equation in $x$, since we define:

\begin{equation}
\begin{array}{rclcrcl}
x_{n}&\equiv& x &,& x_{n-1}&\equiv&x_{n}-h_{-}(x_{n},t)  \\
x_{n+1}&\equiv&x_{n}+h_{+}(x_{n},t) &,&
x_{n+2}&\equiv&x_{n}+h_{+}(x_{n},t)+h_{+}(x_{n+1},t)\,
,
\ldots\\
u_{n}&\equiv& u(x_{n},t) &,& u_{n+k}&\equiv& u(x_{n+k},t)\, .
\end{array}
\end{equation}
At this stage we are not imposing any boundary conditions, so we assume that
equations
(\ref{eq:E=0}) and (\ref{eq:omega=0}) can be shifted arbitrarily to the left and
to
the
right. Thus, eq.(\ref{eq:E=0}) and (\ref{eq:omega=0}) involve any $n_{1}+n_{2}+1$
or
$n_{3}+n_{4}+1$ neighbouring points, respectively.

The fact that (\ref{eq:E=0}) involves only first and second derivatives and that
there
are no derivatives in (\ref{eq:omega=0}) is also for simplicity only. The same
goes
for
the fact that derivatives are evaluated at the reference point $n$ only (i.e. we
do
not
consider terms like $\partial u(x_{n+1},t)/\partial t$).

In order to be able to consider eq.(\ref{eq:E=0}) and (\ref{eq:omega=0}) as a
difference
scheme, we must be able to obtain $x_{n+N},u_{n+N}$ and also $x_{n-M},u_{n-M}$
($N=\max(n_{2},n_{4}), M=\max(n_{1},n_{3})$). In other words, we impose two
conditions:

\begin{equation}
\begin{array}{lcl}
\det \left(\frac{\partial(E,\Omega)}{\partial(x_{n+N},u_{n+N})}\right)\neq 0
&,&
\det \left(\frac{\partial(E,\Omega)}{\partial(x_{n-M},u_{n-M})}\right)\neq 0 \, .
\end{array}
\label{eq:det}
\end{equation}
If necessary, when calculating (\ref{eq:det}) we shift one of the
equations, (\ref{eq:E=0}) or (\ref{eq:omega=0}),
to
the
left or right, so that the same values $n+N$ and $n-M$ figure in both equations.

In general, we do not require that a continuous limit should exist. If it does,
then
eq.(\ref{eq:E=0}) should go into a differential equation in $x$ and $t$ and
eq.(\ref{eq:omega=0}) should go into the identity $0=0$. When taking the
continuous
limit
it is convenient to introduce `discrete derivatives', e.g.

\begin{equation}
\label{eq:d-der}
u,_{x} = \frac{u_{n+1}-u_{n}}{x_{n+1}-x_{n}}, \qquad
u,_{\underline{x}} = \frac{u_{n}-u_{n-1}}{x_{n}-x_{n-1}}, \qquad
u,_{x\bar{x}} = 2\frac{u,_{x}-u,_{\underline{x}}}{x_{n+1}-x_{n-1}}
\end{equation}
etc. In the continuous limit we have
$h_{+}(x_{k},t)\rightarrow 0\, ,\, h_{-}(x_{k},t)\rightarrow 0,\, x_{n+k}\rightarrow
x\, ,\,
u_{k}\rightarrow u(x)$ and the discrete derivatives
go
to
the continuous ones.

A solution of the system (\ref{eq:E=0}), (\ref{eq:omega=0}) will have the form
$x_{n}=\Phi(n,c_{1},\ldots,c_{k})$, $u_{n}=f(x_{n},c_{1},\ldots,c_{k})$ where
$c_{1},\ldots,c_{k}$ are constants needed to satisfy initial conditions and the
functions
$\Phi$ and $f$ are such that (\ref{eq:E=0}) and (\ref{eq:omega=0}) become
identities.

As a clarifying example of eqs.(\ref{eq:E=0}) and (\ref{eq:omega=0}), let us
consider a
three point purely difference scheme, namely

\begin{eqnarray}
E &=& \frac{u_{n+1}-2u_{n}+u_{n-1}}{(x_{n+1}-x_{n})^2}-u_{n}=0,
\label{eq:2-7a} \\*[2ex]
\Omega &=& x_{n+1}-2x_{n}+x_{n-1}=0\, .
\label{eq:2-7b}
\end{eqnarray}
The equation $\Omega=0$ determining the lattice has constant coefficients and
its
solution is $x_{n}=hn+x_{0}$, where $h=h_{+}=h_{-}$ and $x_{0}$ are constants. The
equation $E=0$ on this lattice also has constant coefficients (since we have
$x_{n+1}-x_{n}=h$) and its general solution is
\begin{eqnarray}
u(x_{n})=c_{1}K_{+}^{x_{n}}+c_{2}K_{-}^{x_{n}} &,& K_{\pm}=\left(\frac{2+h^2\pm
h\sqrt{4+h^2}}{2}\right)^{1/h}\, .
\label{eq:2-10}
\end{eqnarray}
In the continuous limit we obtain $E=0\rightarrow u''-u=0\, ,\,
\Omega=0 \rightarrow
0=0\,
,\, u(x)=c_{1}\mathrm{e}^{x}+c_{2}\mathrm{e}^{-x}$. Eq.
(\ref{eq:2-7b})
happens to determine a regular (equally spaced) lattice. Below we shall see
examples
of
other lattices.

\subsection{Symmetries of differential-difference schemes}

Let us consider a one-parameter group of local point transformations of the form

\begin{eqnarray}
\tilde x=\Xi_{\lambda}(x,t,u)\,\, ,\, \, \tilde t=\Gamma_{\lambda}(t)\, \,
,\,\,
\tilde u(\tilde x,\tilde t)=\Phi_{\lambda}(x,t,u)\, .
\label{eq:2-11}
\end{eqnarray}
We shall require that they leave the system of equations (\ref{eq:E=0}),
(\ref{eq:omega=0}) invariant on the solution set of this system. Since we are
interested
in continuous transformations (of discrete systems), we use an infinitesimal
approach
and
write the transformations up to order $\lambda$ as

\begin{eqnarray}
\tilde x&=&x+\lambda\, \xi(x,t,u(x,t))\, ,
\label{eq:2-12} \\
\tilde t&=&t+\lambda\, \tau(t)\, ,
\label{eq:2-13} \\
\tilde u(\tilde x,\tilde t)&=&u(x,t)+\lambda\,
\phi(x,t,u(x,t))\ \ ,\ \ |\lambda|\ll 1\, .
\label{eq:2-14}
\end{eqnarray}
This assumption is quite restrictive. Not only do we consider only
point transformations, but we require that both $t$ and $\tilde t$ are
continuous. No dependence, explicit or implicit, on the discretely
sampled variable $x$ is allowed. Indeed, once the lattice equation is
solved, we get a discrete set of points $\{x_{n}\}$and this would
introduce discrete values $\tilde{t} = \tilde t_{n}$, which we do not
allow. Moreover, the $x$-dependence of $t$, if allowed, remains
unspecified, since the considered equations involve only time
derivatives. This would lead to wrong results, i.e. infinite
dimensional transformation groups that do {\bf not} take solutions
into solutions.

We must now prolong the action of the transformation (\ref{eq:2-12}) to the
prolonged
space. This space includes the derivatives $u_{t}(x,t), u_{tt}(x,t)$, the shifted
points
$x_{\pm}=x_{n\pm 1}, \ldots$ and the function at shifted points
$u_{\pm}=u(x_{\pm},t),\ldots$

It is convenient to express the invariance condition for the system
(\ref{eq:E=0}),
(\ref{eq:omega=0}) in terms of a formalism involving vector fields and their
prolongations. The vector field itself has the form

\begin{equation}
\hat X= \xi(x,t,u)\, \partial_{x}+\tau(t)\, \partial_{t}+\phi(x,t,u)\,
\partial_{u}
\label{eq:2-22}
\end{equation}
with $\xi, \tau$ and $\phi$ the same as in eq.(\ref{eq:2-12})--(\ref{eq:2-14}).
The prolongation of the vector field (\ref{eq:2-22}) acting on the
system
(\ref{eq:E=0}), (\ref{eq:omega=0}) is

\begin{equation}
\mathrm{pr}^{(M+N)}\, \hat X=\hat
X+\sum_{k=n-M}^{n+N}\xi(x_{k},t,u_{k})\partial_{x_{k}}+
\sum_{k=n-M}^{n+N}\phi^{(k)}\partial_{u_{k}}+\phi^{t}\partial_{u_{t}}+
\phi^{tt}\partial_{u_{tt}}
\label{eq:2-23}
\end{equation}
with

\begin{eqnarray}
\phi^{(k)}&=&\phi(x_{k},t,u_{k})
\label{eq:2-24} \\*[2ex]
\phi^{t}&=&D_{t}\phi-(D_{t}\xi)\, u_{x}-(D_{t}\tau)\, u_{t}
\label{eq:2-25} \\*[2ex]
\phi^{tt}&=&D_{t}\phi^{t}-(D_{t}\xi)\, u_{xt}-(D_{t}\tau)\, u_{tt}\, .
\label{eq:2-26}
\end{eqnarray}
Thus the prolongation coefficients $\phi^{t},\phi^{tt}$ are the same as for
differential
equations, the coefficients $\phi^{(k)}$ are as in \cite[\ldots, 27]{10}.

The requirement that the system (\ref{eq:E=0}), (\ref{eq:omega=0}) be invariant
under
the considered one-parameter group translates into the requirement

\begin{equation}
pr\hat X\, E\left |_{E=0\, ,\, \Omega=0}\right.=0\ \ ,\ \ pr\hat X\,
\Omega\left
|_{E=0\, ,\, \Omega=0}
\right.=0\, .
\label{eq:2-27}
\end{equation}
In eq.(\ref{eq:2-27}), once the equations (\ref{eq:E=0}), (\ref{eq:omega=0}) are
taken into account, all involved variables are to be considered as independent.
Eq.(\ref{eq:2-27}) are thus the determining equations for the infinitesimal
coefficients $\xi,\tau$ and $\phi$.

For purely difference equations ($u_{t}$ and $u_{tt}$ absent in (\ref{eq:E=0}))
the
procedure is the following

\begin{enumerate}

\item Extract $u_{n+N}$ and $x_{n+N}$ (or $u_{n-M}$ and $x_{n-M}$) from the
equations
(\ref{eq:E=0}) and (\ref{eq:omega=0})  and substitute into eq.(\ref{eq:2-27}).
This
provides us with two functional equations for $\xi,\tau$ and $\phi$.

\item Assuming an analytical dependence of $\xi,\tau$ and $\phi$ on their own
variables, we convert these two equations into differential equations by
differentiating them with respect to appropriately chosen variables
$u_{n+k},\, x_{n+k}$. Use the fact that the coefficients $\xi,\tau$ and $\phi$
depend
on
$x$ and $u$ evaluated at one point only to simplify the equations. Differentiate
sufficiently many times to obtain differential equations that we can integrate.

\item Solve the differential equations, substitute back into the two original
functional equations and solve them.

\end{enumerate}

For differential-difference equations, we solve for the highest derivative (in our
case $u_{tt}$) and for either $x_{n+N}$, or $u_{n+N}$ (or $x_{n-M}$ or $u_{n-M}$)
and
substitute into eq.(\ref{eq:2-27}). In this case, the determining equation will be
a polynomial expression in the
derivatives of $u$ with respect to $t$ (in our case $u_{t}$ only) and all their
coefficients must vanish. For the remaining terms, which depend on shifted
variables,
we proceed as in the case of purely difference equations.

\section{Examples of symmetries of difference equations}

We shall give several examples of the calculation of symmetries acting on
difference
schemes. They will involve either three or four points on a lattice. Equations
(\ref{eq:E=0}) and (\ref{eq:omega=0}) simplify to

\begin{eqnarray}
E(x,x_{-},x_{+},x_{++},u,u_{-},u_{+},u_{++})=0
\label{eq:3-1} \\*[2ex]
\Omega(x,x_{-},x_{+},x_{++},u,u_{-},u_{+},u_{++})=0
\label{eq:3-2}
\end{eqnarray}
for a four point scheme. A three point scheme is obtained if $E$ and $\Omega$ are
independent of $x_{++}$ and $u_{++}$. Here $x=x_{n}$ is the reference point and
$x_{-}=x_{n-1},\, x_{+}=x_{n+1},\, x_{++}=x_{n+2}$ and similarly for $u$.

The prolongation (\ref{eq:2-23}) of the vector field simplifies to

\begin{equation}
\begin{array}{rcl}
pr \hat X &=&\xi(x,u)\partial_{x}+\phi(x,u)
\partial_{u}+\xi(x_{-},u_{-})\partial_{x_{-}}+\xi(x_{+},u_{+})
\partial_{x_{+}}+\phi(x_{-},u_{-})\partial_{u_{-}} \\*[2ex] &&
+\xi(x_{++},u_{++})\partial_{x_{++}}+\phi(x_{+},u_{+})
\partial_{u_{+}}+\phi(x_{++},u_{++})
\partial_{u_{++}}\ .
\label{eq:3-3}
\end{array}
\end{equation}
(for three point schemes we drop the $x_{++},\, u_{++}$ terms).

A symmetry classification of three point schemes was provided in a recent article
\cite{35}. Here we solve a different problem. The equations and lattices are given
and we determine their symmetries.

\subsection{Polynomial nonlinearity on a uniform lattice}

Let us consider the nonlinear ordinary differential equation

\begin{equation}
u_{xx}-u^{N}=0\ \ ,\ \ N\neq 0,1\, .
\label{eq:3-4}
\end{equation}
A straightforward calculation shows that for $N\neq -3$ eq.(\ref{eq:3-4})  is
invariant
under a two-dimensional Lie group, the Lie algebra of which is spanned by

\begin{equation}
\hat P=\partial_{x}\ \ ,\ \ \hat D=(N-1)x\partial_{x}-2u\partial_{u}\, .
\label{eq:3-5}
\end{equation}
For $N=-3$ the symmetry algebra is $sl(2,\Re)$ with a basis

\begin{equation}
\hat P=\partial_{x}\ \ ,\ \ \hat D=2x\partial_{x}+u\partial_{u}\ \ ,\ \
\hat C=x^2\partial_{x}+xu\partial_{u}\, .
\label{eq:3-6}
\end{equation}

A natural way to discretize eq.(\ref{eq:3-4}) is to use a uniform lattice and put

\begin{eqnarray}
E&=&\frac{u_{+}-2u+u_{-}}{(x_{+}-x)^{2}}-u^{N}=0
\label{eq:3-7a} \\*[2ex]
\Omega&=&x_{+}-2x+x_{-}=0\, .
\label{eq:3-7b}
\end{eqnarray}
Let us now apply the symmetry algorithm (\ref{eq:2-27}). The condition $pr X\,
\Omega=0$ for $E=0,\, \Omega=0$ implies

\begin{equation}
\xi(2x-x_{-},(x-x_{-})^{2}u^{N}+2u-u_{-})-2\xi(x,u)+\xi(x_{-},u_{-})=0\, .
\label{eq:3-8}
\end{equation}
Differentiating first by $\partial_{u_{-}}$, then by $\partial_{u}$ we obtain

\begin{eqnarray}
-\xi_{u_{+}}(2x-x_{-},(x-x_{-})^{2}u^{N}+2u-u_{-})+\xi_{u_{-}}(x_{-},u_{-})&=&0
\label{eq:3-9} \\*[2ex]
\left[N(x-x_{-})^2\, u^{N-1}+2\right]\,
\xi_{u_{+}u_{+}}(2x-x_{-},(x-x_{-})^{2}u^{N}+2u-u_{-})&=&0\, .
\label{eq:3-10}
\end{eqnarray}
Eq.(\ref{eq:3-10}) implies that $\xi$ is linear in $u$
\begin{equation}
\xi(x,u)=a(x)u+b(x)\, .
\label{eq:3-11}
\end{equation}
Eq.(\ref{eq:3-9}) reduces to $a(x_{+})=a(x)$, i.e. $a$ is a constant. Substituing
these results into eq.(\ref{eq:3-8}) we obtain

\begin{equation}
a\, [u_{+}-2u+u_{-}]+b(x_{+})-2b(x)+b(x_{-})=0\, .
\label{eq:3-12}
\end{equation}
This implies $a=0$ and

\begin{equation}
b(x_{+})-2b(x)+b(x_{-})=0\, .
\label{eq:3-13}
\end{equation}
Differentiating successively with respect to $x$ and $x_{-}$ we find
$b_{x_{+}x_{+}}(x_{+})=0$, i.e.

\begin{equation}
b(x)=b_{1}x+b_{0}\, .
\label{eq:3-14}
\end{equation}

Thus, the invariance of eq.(\ref{eq:3-7b}) implies $\xi=b_{1}x+b_{0}$ with
$b_{1},\,
b_{0}$ constants. The function $\phi(x,u)$ is restricted by the requirement $pr
X\,
E=0$ for $E=0,\, \Omega=0$. This invariance condition is given by

\begin{equation}
\begin{array}{l}
\phi(2x-x_{-},(x-x_{-})^{2}u^{N}+2u-u_{-})-2\phi(x,u)+\phi(x_{-},u_{-}) \\*[2ex]
-(x-x_{-})^{2}[N\phi(x,u)u^{N-1}+2b_{1}u^N]=0\, .
\end{array}
\label{eq:3-15}
\end{equation}
We successively differentiate this equation with respect to $u_{-}$ and $u$ and we
obtain

\begin{eqnarray}
-\phi_{u_{+}}(x_{+},u_{+})+\phi_{u_{-}}(x_{-},u_{-})&=&0
\label{eq:3-16a} \\*[2ex]
\phi_{u_{+}u_{+}}(x_{+},u_{+})&=&0\, .
\label{eq:3-16b}
\end{eqnarray}
These two equations require that $\phi=\phi_{1}u+\phi_{0}(x)$ with $\phi_{1}$ a
constant. Substituing back into eq.(\ref{eq:3-15}) we obtain the remaining
determining equation

\begin{equation}
\phi_{0}(2x-x_{-})-2\phi_{0}(x)+\phi_{0}(x_{-})-(x-x_{-})^{2}[(N-1)\phi_{1}+2b_{1}
] u^{N}
-N(x-x_{-})^{2}\phi_{0}u^{N-1}=0
\label{eq:3-17}
\end{equation}
Since we have $N\neq0,1$ eq.(\ref{eq:3-17}) implies $\phi_{0}(x)=0$ and
$\phi_{1}(1-N)=2b_{1}$. Finally, we obtain the symmetry algebra of the difference
system (\ref{eq:3-7a}), (\ref{eq:3-7b}). It is $2$-dimensional and coincides with
the
algebra (\ref{eq:3-5}) of the differential equation (\ref{eq:3-4}), the continuous
limit of eq.(\ref{eq:3-7a}).

Notice that the case $N=-3$ is not distinguished from the generic case. As a
matter
of fact, no difference equation on a uniform lattice can be invariant under the
$SL(2,\Re)$ group corresponding to the algebra (\ref{eq:3-6}). A basis for
the
difference invariants of this algebra in the space
$\{x,x_{-},x_{+},u,u_{-},u_{+}\}$
is

\begin{equation}
\rho_{1}=\frac{h_{-}u_{+}}{(h_{+}+h_{-})u}\ \ ,\ \
\rho_{2}=\frac{h_{+}u_{-}}{(h_{+}+h_{-})u}\ \ ,\ \
\rho_{3}=\frac{h_{+}h_{-}}{(h_{+}+h_{-})u^2}
\label{eq:3-18}
\end{equation}
where $h_{+}$ and $h_{-}$ are defined as $h_{+}=x_{+}-x,\,
h_{-}=x-x_{-}$, and no function of $x$, $x_{+}$ and $x_{-}$ alone can
be set equal to a constant. An $SL(2,\Re)$ invariant scheme must
be
constructed out of these invariants. For instance, an invariant scheme
approximating
eq.(\ref{eq:3-4}) for $N=-3$ is

\begin{eqnarray}
\frac{h_{-}(u_{+}-u)-h_{+}(u-u_{-})}{h_{+}h_{-}(h_{+}+h_{-})}&=&\frac{2h_{+}h_{-}}
{(h
_{+}+h_{-})^2}\frac{1}{u^3}\ \ ,\ \ h_{-}u_{+}=h_{+}u_{-}\, .
\label{eq:3-19}
\end{eqnarray}

\subsection{Discrete versions of linear second order equations}

\subsubsection{Discretization of $u_{xx}=u$}

Consider the ordinary differential equation

\begin{equation}
u_{xx}=u.
\label{eq:3-20}
\end{equation}
Like every second order linear ODE, it is invariant under $SL(3,\Re)$ with
the
Lie algebra realized in this case by the vector fields

\begin{equation}
\begin{array}{c}
\hat X_{1}=\partial_{x}\ ,\ \hat X_{2}=u\partial_{u}\ ,\
\hat X_{3}= e^{x}\partial_{u}\ ,\ \hat X_{4}= e^{-x}\partial_{u}\ ,\
\hat X_{5}= e^{2x}(\partial_{x}+u \partial_{u}) ,\\*[2ex]
\hat X_{6}=u e^{x}(\partial_{x}+ u \partial_{u})\ ,\
\hat X_{7}= e^{-2x}(\partial_{x}-u \partial_{u})\ ,\
\hat X_{8}=u e^{-x}(\partial_{x}- u \partial_{u})\, .
\end{array}
\label{eq:3-21}
\end{equation}

A very straightford discretization of eq.(\ref{eq:3-20}) on a uniform lattice is

\begin{eqnarray}
\frac{u_{+}-2u+u_{-}}{(x_{+}-x)^2}&=&u ,
\label{eq:3-22a} \\*[2ex]
x_{+}-2x+x_{-}&=&0\, .
\label{eq:3-22b}
\end{eqnarray}

Applying the same procedure to the system (\ref{eq:3-22a}), (\ref{eq:3-22b}) that
was
applied to the system (\ref{eq:3-7a}), (\ref{eq:3-7b}) (with $N\neq0,1$), we again
obtain a $2$-dimensional symmetry algebra

\begin{equation}
\hat P=\partial_{x}\ \ ,\ \ \hat D=u\partial_{u}\, .
\label{eq:3-23}
\end{equation}
At first glance the absence of symmetries of the form $\phi(x)\partial_{u}$,
representing the linear superposition principle, seems surprising. However, viewed
as
a system of two equations, the system (\ref{eq:3-22a}), (\ref{eq:3-22b}) is really
nonlinear. Eq.(\ref{eq:3-22b}) defines a uniform lattice with an arbitrary step
$h=x_{+}-x=x-x_{-}$, where the step $h$ can be scaled by a dilatation of $x$.

An alternative approach to the system (\ref{eq:3-22a}), (\ref{eq:3-22b}) is to
first
integrate eq. (\ref{eq:3-22b}) once, thus fixing the step on the $x$-axis. The
system
(\ref{eq:3-22a}), (\ref{eq:3-22b}) is then replaced by the equation
\begin{equation}
\frac{u_{+}-2u+u_{-}}{h^2}=u\ ,
\label{eq:3-24}
\end{equation}
where $h=x_{+}-x=x-x_{-}$ is a fixed (non-scalable) constant. The symmetry algorithm
described in
Section~2 and applied in Section~3.1 yields a four-dimensional symmetry algebra

\begin{equation}
\hat P=\partial_{x}\ \ ,\ \ \hat D=u\partial_{u}\ \ ,\ \ \hat
S_{1}=K_{+}^{x}\partial_{u}\ \ ,\ \
\hat S_{2}=K_{-}^{x}\partial_{u}
\label{eq:3-25}
\end{equation}
with $K_{\pm}$ as in eq.(\ref{eq:2-10}). The symmetries $\hat S_{1},\, \hat S_{2}$
represent
the linear superposition formula for the linear system (\ref{eq:3-24}).

We mention that eq.(\ref{eq:3-20}) (and any linear ODE) can be discretized in a
manner
that exactly preserves all of its solutions. To do this we must preserve a
subalgebra
of the symmetry algebra of the ODE, containing the elements corresponding to the
linear superposition formula. In our case these are $\hat X_{3}$ and $\hat X_{4}$
of
eq.(\ref{eq:3-21}). Let us consider the subalgebra $\{\hat X_{1},\ldots,\hat
X_{6}\}$. Its second order discrete prolongation allows no invariants. It does
however allow an invariant manifold, namely

\begin{equation}
I=u e^{-x}( e^{-2x_{+}}- e^{-2x_{-}})+u_{+} e^{-x_{+}}( e^{-2x_{-}}- e^{-2x}
)+u
_{-} e^{-x_{-}}( e^{-2x}- e^{-2x_{+}})=0\, .
\label{eq:3-26}
\end{equation}
The expression

\begin{equation}
S=\frac{ e^{-2x}- e^{-2x_{-}}}{ e^{-2x_{+}}- e^{-2x}}
\label{eq:3-27}
\end{equation}
is an invariant on the manifold (\ref{eq:3-26}).

Indeed, we have

\begin{equation}
\begin{array}{c}
(\hat X_{1}+3\hat X_{2})\, I=0\ \ ,\ \ \hat X_{3}\, I=\hat X_{4}\, I=\hat X_{5}\,
I=\hat X_{6}\, I=0\ \ ,\ \ \hat X_{2}\, I=I ,  \\*[2ex]
\hat X_{i}\, S=0\ \ (i=1,\ldots,5)\ \ ,\ \ \hat X_{6}\, S=\frac {2\,
I}{( e^{-2x_{+}}- e^{-2x})^2}\, ,
\label{eq:3-28}
\end{array}
\end{equation}
so that we have

\begin{equation}
\hat X_{i}\, I\left|_{I=0}=\right.0\ \ ,\ \ \hat X_{i}\, S\left|_{I=0}=\right.0\ \
,\ \
i=1,\ldots,6\, .
\end{equation}
A uniform lattice, to first order in $h$ and an equation with (\ref{eq:3-20}) as
its
continuous limit, is obtained by putting

\begin{equation}
S=1\ \ \ ,\ \ \ \frac{ e^{3x}\, I}{2h^3}=0\, .
\label{eq:3-29}
\end{equation}
Eq.(\ref{eq:3-29}), or $I=0$, has $u= e^{x}$ and $u= e^{-x}$ as solutions and
the
general solution is

\begin{equation}
u=c_{1} e^{x}+c_{2} e^{-x}\, ,
\label{eq:3-30}
\end{equation}
just as in the continuous case (\ref{eq:3-20}).

To check this, let us solve the system $S=1$, $I=0$ directly, with $I$
and $S$ given in eq. (\ref{eq:3-26}) and (\ref{eq:3-27}),
respectively. We linearize $S=1$ by a change of variables and obtain:
\begin{equation}
z=e^{-2x}\ \ ,\ \ z_{+} - 2 z + z_{-} = 0.
\label{eq:3-30a}
\end{equation}
The solution is:
\begin{equation}
z_{n} = c_{3} n + c_{4} , \, \ \ x_{n} = -\frac{1}{2} \ln ( c_{3} n +
c_{4} ) ,
\label{eq:3-30b}
\end{equation}
so that the lattice in $x$ is logarithmic ( $c_{3}$ and $c_{4}$ are
integration constants).  On this lattice eq.(\ref{eq:3-26}) reduces
to
\begin{equation}
2u \sqrt{c_{3} n + c_{4}} - u_{+} \sqrt{ c_{3} (n+1)
 + c_{4}} - u_{-} \sqrt{ c_{3} (n-1) + c_{4}}
=0.
\label{eq:3-30c}
\end{equation}
To solve this linear equation we put $u(x)=e^{x}f(x)$ or, on the
lattice
\begin{equation}
u(x_{n}) = \frac{1}{\sqrt{ c_{3} n + c_{4}}} f(x_{n}),
\label{eq:3-30d}
\end{equation}
so that $f(x)$ satisfies
\begin{equation}
f(x_{+}) - 2 f(x) + f(x_{-}) = 0.
\label{eq:3-30e}
\end{equation}
We write the general solution of eq.(\ref{eq:3-30e}) as
\begin{equation}
f(x_{n}) = f\circ x(n) = A\, n + B\ .
\label{eq:3-30f}
\end{equation}
By rewriting $A$ and $B$ in terms of the new integration constants $c_{1}$ and $c_{2}$,
i.e. by putting $A = c_{2}\, c_{3}$ and $B = c_{2}\, c_{4} + c_{1}$, we obtain the
general solution of the system (\ref{eq:3-29}) as
\begin{equation}
u = \frac{c_{1}}{\sqrt{ c_{3} n + c_{4}}} +
c_{2} \sqrt{ c_{3} n + c_{4}} = c_{1} \me^{x} + c_{2} \me^{-x}
\label{eq:3-30g}
\end{equation}
in full agreement with eq.(\ref{eq:3-30}).

\subsubsection{Discrete version of $u_{xx}=1$}

Let us consider the simplest 3 point difference scheme for the ODE $u_{xx}=1$

\begin{equation}
\frac{u_{+}-2u+u_{-}}{(x_{+}-x)^2}=1\ \ ,\ \ x_{+}-2x+x_{-}=0\, .
\label{eq:3-31}
\end{equation}
Applying the prolonged vector field to these equations and eliminating $x_{+}$
and
$u_{+}$, we obtain two equations

\begin{eqnarray}
\xi(2x-x_{-},(x-x_{-})^{2}+2u-u_{-})-2\xi(x,u)+\xi(x_{-},u_{-})=0 ,
\label{eq:3-32} \\*[2ex]
\phi(2x-x_{-},(x-x_{-})^{2}+2u-u_{-})-2\phi(x,u)+\phi(x_{-},u_{-})=
\nonumber \\*[2ex]
2(x-x_{-})\left[\xi(2x-x_{-},(x-x_{-})^{2}+2u-u_{-})-\xi(x,u)\right]\, .
\label{eq:3-33}
\end{eqnarray}
We first concentrate on eq.(\ref{eq:3-32}). Taking the second derivative with
respect
to $u$ and $u_{-}$ we find that $\xi$ is linear in $u$. Substituing back into
(\ref{eq:3-32}) and differentiating with respect to $x$ and $x_{-}$ we find

\begin{equation}
\xi(x,u)=\alpha(u-\frac{x^2}{2})+\beta_{1}x+\beta_{0}
\label{eq:3-34}
\end{equation}
where $\alpha,\, \beta_{1}$ and $\beta_{0}$ are constants. Substituing $\xi$ into
eq.(\ref{eq:3-33}) and solving for $\phi$ in a similar manner, we obtain:

\begin{equation}
\phi(x,u)=\alpha(xu-\frac{x^3}{2})+c(u-\frac{x^2}{2})+\beta_{1}x^2+
\beta_{2}x+\beta_{3} .
\label{eq:3-35}
\end{equation}

Finally, a basis for the symmetry algebra of the system (\ref{eq:3-31}) is

\begin{equation}
\begin{array}{c}
\hat X_{1}=\partial_{x}\ ,\ \hat X_{2}=\partial_{u}\ ,\ \hat X_{3}=x\partial_{u}\
,\
\hat X_{4}=x\partial_{x}+x^2\partial_{u} ,\\*[2ex]
\hat X_{5}=(u-\frac{x^2}{2})\partial_{u}\ ,\ \hat
X_{6}=(u-\frac{x^2}{2})\partial_{x}+(u-\frac{x^2}{2})x\partial_{u}\, .
\end{array}
\label{eq:3-36}
\end{equation}
It is easy to check that this Lie algebra is isomorphic to the
general affine Lie algebra
$gaff(2,\Re)$. This is the symmetry algebra of the scheme \cite{35}

\begin{equation}
w_{+}-2w+w_{-}=0\ \ ,\ \ t_{+}-2t+t_{-}=0\, .
\label{eq:3-37}
\end{equation}
Indeed the system (\ref{eq:3-31}) is transformed into (\ref{eq:3-37}) by
putting

\begin{equation}
u=w+\frac{t^2}{2}\ \ ,\ \ x=t\, .
\label{eq:3-38}
\end{equation}

\subsection{Discrete versions of the equation $u_{xxx}=0$}

The symmetry algebra of the ODE $u_{xxx}=0$ is $7$-dimensional. A basis for this
algebra is

\begin{equation}
\begin{array}{c}
\hat X_{1}=\partial_{x}\ ,\
\hat X_{2}=\partial_{u}\ ,\
\hat X_{3}=x\partial_{x}\ ,\
\hat X_{4}=u\partial_{u}\ ,\
\hat X_{5}=x\partial_{u} , \\*[2ex]
\hat X_{6}=x^2\partial_{u}\ ,\
\hat X_{7}=x^2\partial_{x}+2xu\partial_{u}\, .
\end{array}
\label{eq:3-39}
\end{equation}
The generators $\hat X_{2},\, \hat X_{5},\, \hat X_{6}$ correspond to the linear
superposition principle. We can add $u=c_{2}x^2+c_{1}x+c_{0}$ to any solution and
indeed, this itself is the general solution.

Let us now consider discretizations of this ODE.

\subsubsection{Discretization on a uniform lattice}

We consider the system

\begin{eqnarray}
E&=&u_{++}-3u_{+}+3u-u_{-}=0 ,
\label{eq:3-40a} \\*[2ex]
\Omega_{1}&=&x_{+}-2x+x_{-}=0 .
\label{eq:3-40b}
\end{eqnarray}
The lattice is uniform, since the general solution of (\ref{eq:3-40b}) is
$x_{n}=hn+x_{0}$ with $h$ and $x_{0}$ constants. Eq.(\ref{eq:3-40b}) must be
shifted
once to the right to obtain $x_{++}$.

The prolonged vector fields have the form (\ref{eq:3-3}). We apply the same method
as
in Section~3.2. to obtain the symmetry algebra of the system
(\ref{eq:3-40a}), (\ref{eq:3-40b}). The result
is a
$6$-dimensional Lie algebra generated by $\{\hat X_{1}\, ,\, \hat X_{2}\, ,\,
\hat X_{3}\, ,\, \hat X_{4}\, ,\, \hat X_{5}\, ,\, \hat X_{6}\}$ of
eq.(\ref{eq:3-39}). The system hence has exactly the same solutions as the ODE
$u_{xxx}=0$, however the lattice is not invariant under the projective
transformations generated by $\hat X_{7}$.

\subsubsection{Discretization on a four point lattice}

We take the equation (\ref{eq:3-40a}) on the lattice

\begin{equation}
\Omega_{2}=x_{++}-3x_{+}+3x-x_{-}=0\, .
\label{eq:3-41}
\end{equation}
The lattice given by equation (\ref{eq:3-41}) is not uniform but satisfies
$x_{n}=L_{2}n^2+L_{1}n+L_{0}$, where
$L_{i}$
are constants. We assume $L_{2}\neq 0$, otherwise the lattice is the same as for
$\Omega_{1}=0$.

The symmetry algebra in this case is given by

\begin{equation}
\{\hat X_{1}\, ,\, \hat X_{2}\, ,\, \hat X_{3}\, ,\, \hat X_{4}\, ,\, \hat X_{5}\,
,\, \hat
Y=u\partial_{x}\}
\label{eq:3-43}
\end{equation}
with $\hat X_{1},\ldots,\hat X_{5}$ as in eq.(\ref{eq:3-39}). Thus $\hat X_{6}$ of
(\ref{eq:3-39}) is absent. This reflects the fact that $u=x^2$ is not an exact
solution on the lattice $\Omega_{2}=0$. Indeed, if we take $L_{2}=1$ and
$L_{1}=L_{0}=0$ in eq.(\ref{eq:3-39}) we have $u=n^4$ which would solve a fourth
order equation, not however equation (\ref{eq:3-40a}).

\subsubsection{Discretization preserving the entire symmetry group}

The third prolongation of the algebra (\ref{eq:3-39}) acts on an 8-dimensional
space
with coordinates $(x\, ,\, x_{+}\, ,\, x_{++}\, ,\, x_{-}\, ,\, u\, ,\, u_{+}\,
,\, u_{++}\,
,\, u_{-})$. If
the
7 prolonged fields are linearly independent, they will allow only one invariant.
This
invariant can be calculated directly. It lies entirely in the subspace $\{x\, ,\,
x_{+}\, ,\, x_{++}\, ,\, x_{-}\}$ and is given by the anharmonic ratio of four
points,
namely

\begin{equation}
\frac{(x_{++}-x)(x_{+}-x_{-})}{(x-x_{-})(x_{++}-x_{+})}=K\, .
\label{eq:3-44}
\end{equation}
This is the invariant of the projective action of $sl(2,\Re)$ on the real
line
$\Re$, given by the $\partial_{x}$ part of the subalgebra $\{\hat X_{1}\ ,\
\hat X_{3}\ ,\ \hat X_{7}\}$ of the algebra (\ref{eq:3-39}). Eq.(\ref{eq:3-44})
provides us with a lattice. The invariant equation is obtained by requiring that
the
third prolongation of $(\hat X_{1},\ldots,\hat X_{7})$ be linearly connected on
some
manifold. This manifold is given by the condition

\begin{equation}
\begin{array}{lcl}
I=&-&(u_{+}-u)(x_{++}-x)(x-x_{-})(x_{++}-x_{-}) \\*[2ex]
&+&(u_{++}-u)(x_{+}-x)(x-x_{-})(x_{+}-x_{-}) \\*[2ex]
&+&(u-u_{-})(x_{+}-x)(x_{++}-x)(x_{++}-x_{+})=0\, .
\end{array}
\label{eq:3-45}
\end{equation}
It is easy to check that $I$ is indeed invariant, i.e.

\begin{equation}
\begin{array}{rcl}
pr^{(3)}\hat X_{i}\, I\left|_{I=0}\right.=0 &,& i=1,\ldots,7\, .
\end{array}
\end{equation}
Finally, a difference scheme, invariant under the group generated by the algebra
(\ref{eq:3-39}), having $u_{xxx}=0$ as a continuous limit is given by

\begin{equation}
u,_{\underline{x}x\bar{x}}=\frac{6\,
I}{(x_{++}-x_{-})(x_{++}-x)(x_{++}-x_{+})(x_{+}-x_{-})(x_{+}-x)(x-x_{-})}=0
\label{eq:3-46}
\end{equation}
and eq.(\ref{eq:3-41}).

We define discrete derivatives as

\begin{equation}
\begin{array}{rclcrclcrcl}
u,_{x}&=&\frac{u_{+}-u}{x_{+}-x}
&,& u,_{\bar{x}}&=&\frac{u_{++}-u_{+}}{x_{++}-x_{+}}
&,& u,_{\underline{x}}&=&\frac{u-u_{-}}{x-x_{-}} , \\*[2ex]
u,_{x\bar{x}}&=&2\frac{u_{\bar{x}}-u_{x}}{x_{++}-x} &,&
u,_{x\underline{x}}&=&2\frac{u_{x}-u_{\underline{x}}}{x_{+}-x_{-}} , \\*[2ex]
u,_{\underline{x}x\bar{x}}&=&3\frac{u_{x\bar{x}}-u_{x\underline{x}}}{x_{++}-x_{-}}
\, .
\end{array}
\end{equation}

Any four solutions of a Riccati equation satisfy eq.(\ref{eq:3-44}) and we use this
fact to solve this equation. Indeed, consider e.g. the Riccati equation

\begin{equation}
\dot{x}=Ax^2+Bx+C \ \ ,\ \ B^2-4AC>0
\label{eq:3-48}
\end{equation}
where $A,\, B$ and $C$ are real constants and $A\neq 0$. The general solution of
eq.(\ref{eq:3-48}) is

\begin{equation}
x=\frac{x_{1}+x_{2}\, \omega\,  e^{A(x_{1}-x_{2})t}}{1-\omega\,
 e^{A(x_{1}-x_{2})t}}\ \ ,\ \ x_{1,2}=\frac{-B\pm \sqrt{B^2-4AC}}{2A}\, .
\label{eq:3-49}
\end{equation}
Let us take $\omega=n,\, x_{1}=\alpha,\, x_{2}=\beta$ and
$e^{A(x_{1}-x_{2})t}=\gamma$. A solution of eq.(\ref{eq:3-49}) is

\begin{equation}
x\equiv x(n)=\frac{\alpha n +\beta}{\gamma n + \delta}\ \ ,\ \ \alpha,\, \beta,\,
\gamma, \, \delta =const. \, , \, \alpha \delta - \beta \gamma = 1.
\label{eq:3-50}
\end{equation}
Substituting into eq.(\ref{eq:3-44}) we find $K=4$. The value $K=4$ is also
required
to obtain the correct continuous limit. Indeed, putting
$x_{+}-x=\epsilon\sigma_{1},\,
x-x_{-}=\epsilon\sigma_{2},\, x_{++}-x_{+}=\epsilon\sigma_{3},\, \sigma_{i}\in
\Re$ and $\epsilon\rightarrow 0$ we have

\begin{equation}
\frac{\epsilon^2(\sigma_{1}+\sigma_{3})(\sigma_{1}+\sigma_{2})}{\epsilon^{2}
\sigma_{2}\sigma_{3}}=K
\end{equation}
and for $\sigma_{1}=\sigma_{2}=\sigma_{3}$ we have $K=4$ and also
$u,_{x}\rightarrow
u',\, u,_{\bar{x}}\rightarrow u',\,  u,_{\underline{x}}\rightarrow u',\,
u,_{x\bar{x}}\rightarrow u'',\, u,_{\underline{x}x}\rightarrow u'',\,
u,_{\underline{x}x\bar{x}}\rightarrow u'''$, where the primes denote
(continuous)
derivatives.

Plots of $x(n)$ for lattices (\ref{eq:3-40b}), (\ref{eq:3-41}) and (\ref{eq:3-50})
are shown on
Figure 1,2 and 3, respectively. The expression (\ref{eq:3-50}) is singular for
$\gamma=\delta/n$, so
such values of $\gamma$ are to be avoided.

\begin{figure}
  \begin{center}
 \includegraphics*[scale=0.5]{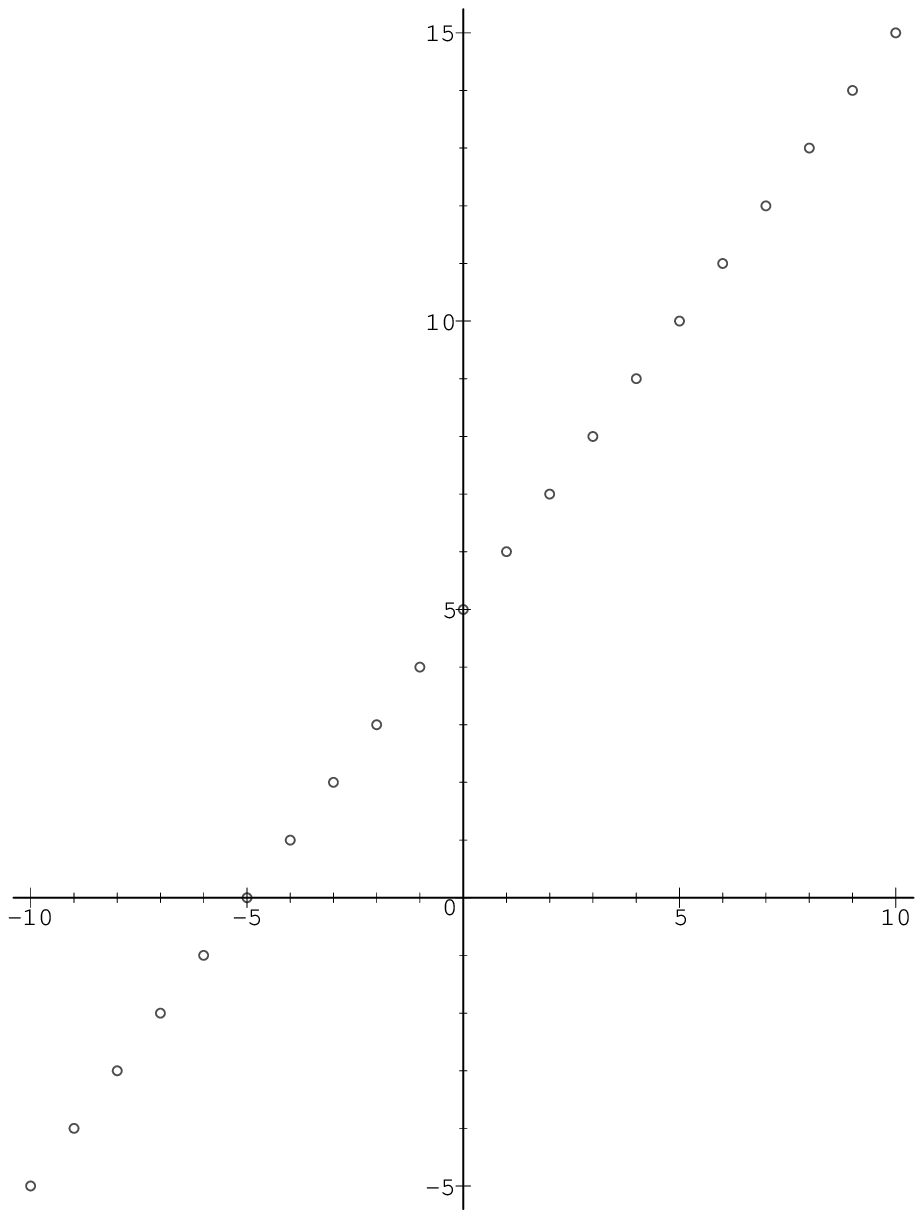}

    \caption{Variable $x$ as a function of $n$ for the lattice (\ref{eq:3-40b})
$x_{n}=hn+x_{0}$ \qquad ($h=1\, ,\, x_{0}=5$)}
    \label{fig:1}
  \end{center}
\end{figure}

\begin{figure}
  \begin{center}
\includegraphics*[scale=0.6]{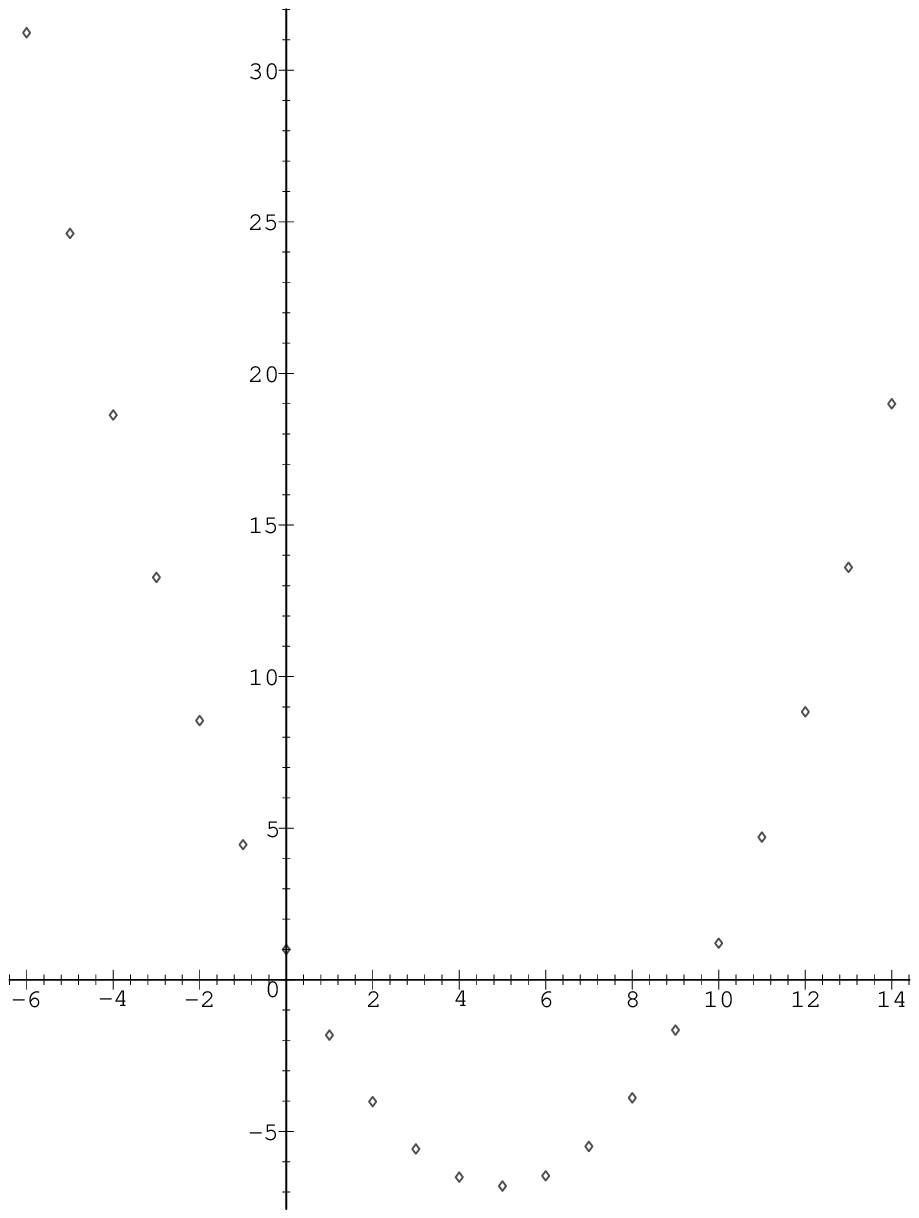}

    \caption{Variable $x$ as a function of $n$ for the lattice (\ref{eq:3-41})
$x_{n}=L_{2}n^2+L_{1}n+L_{0}$ \qquad ($L_{2}=1/\sqrt{10}\, ,\, L_{1}=-\pi\, ,\,
L_{0}=1$)}
    \label{fig:2}
  \end{center}
\end{figure}

\begin{figure}
  \begin{center}
   \includegraphics*[scale=0.6]{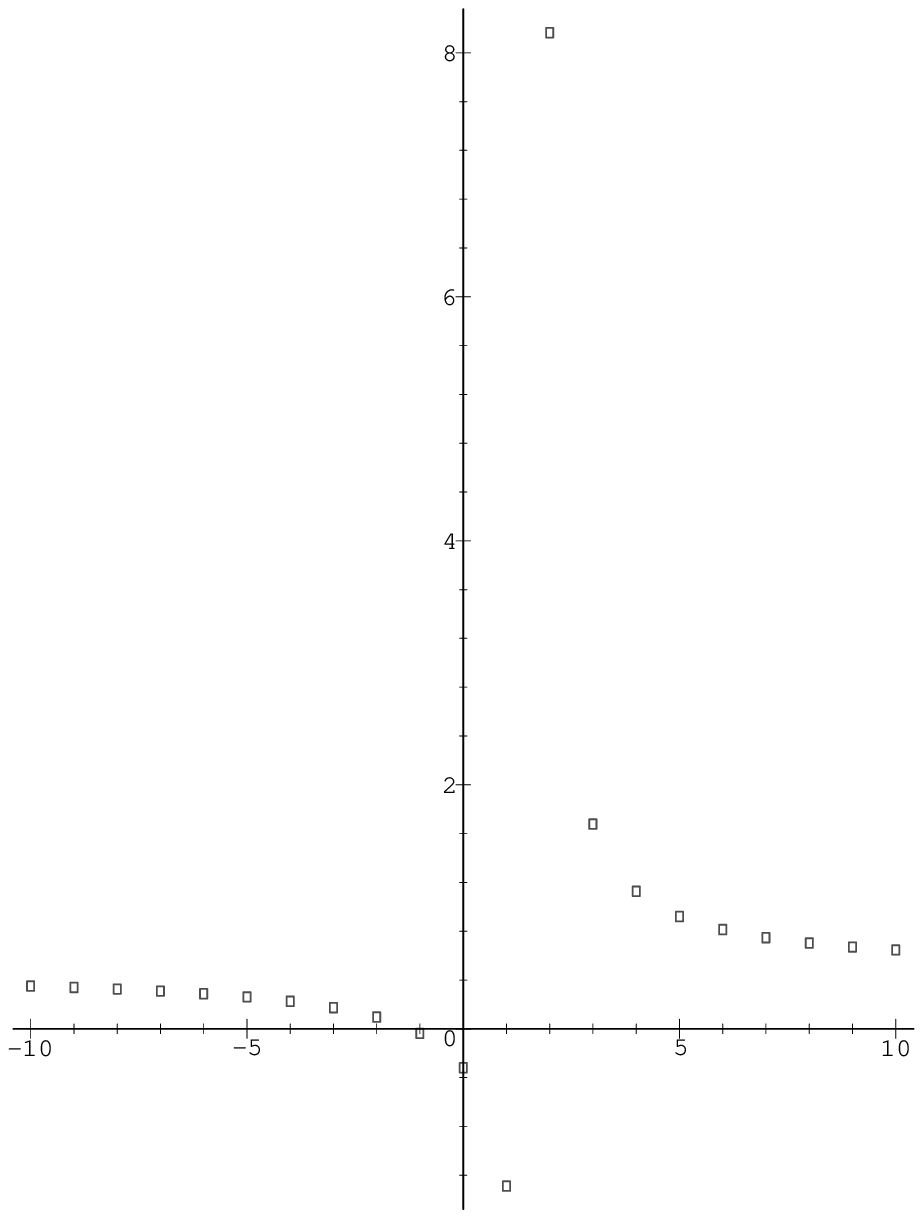}

    \caption{Variable $x$ as a function of $n$ for the lattice (\ref{eq:3-50})
$x_{n}=(\alpha n +\beta)(\gamma n + \delta)^{-1}$ \qquad ($\alpha= \sqrt{2}$ ,
$\beta=-\sqrt{3}$ ,
$\gamma=3$,  $\delta= - \sqrt{3} \pi$)}
    \label{fig:3}
  \end{center}
\end{figure}

\section{Examples for differential-difference equations}

In this section we shall need the complete formalism of Section~2, in particular
the
vector field prolongation (\ref{eq:2-23}),...,(\ref{eq:2-26}).

\subsection{Symmetries of the discrete Volterra equation}

The discrete Volterra equation \cite{17} on a uniform lattice is represented by
the
two
equations

\begin{eqnarray}
E &\equiv& u_{t}+u\, \frac{u_{+}-u_{-}}{x_{+}-x_{-}}=0 ,
\label{eq:4-1a} \\*[2ex]
\Omega &\equiv& x_{+}-2x+x_{-}=0 ,
\label{eq:4-1b}
\end{eqnarray}
where $t$ is a continuous variable, $u=u(x,t)$ and $u_{t}=\partial u/\partial t$.
The
Volterra equation is integrable \cite{17} but we make no use of that here.

The invariance condition for the lattice (\ref{eq:4-1b}) is

\begin{equation}
\xi(2x-x_{-},t,u_{+})-2\xi(x,t,u)+\xi(x_{-},t,u_{-})=0\, .
\label{eq:4-2}
\end{equation}
Contrary to the cases in Section~3, the values $u_{+},\, u$ and $u_{-}$ in
eq.(\ref{eq:4-2}) are independent, since the equation $E=0$ involves $u_{t}$ (in
addition to $u_{+},\, u$ and $u_{-}$). Differentiating eq.(\ref{eq:4-2}) with
respect
to e.g. $u$ we obtain $\xi_{u}=0$. Differentiating with respect to $x_{-}$ and
then
$x$, we obtain $\xi_{x_{+}x_{+}}(x_{+},t)=0$. The function $\xi(x,t,u)$ hence
reduces
to

\begin{equation}
\xi=a(t)x+b(t)
\label{eq:4-3}
\end{equation}
with $a(t)$ and $b(t)$ so far arbitrary functions of $t$.

Invariance of the equation (\ref{eq:4-1a}) implies:

\begin{equation}
\phi^{t}+\phi\frac{u_{+}-u_{-}}{x_{+}-x_{-}}+\frac{u}{x_{+}-x}(\phi^{(+)}-\phi^{(-
)})
-\frac{u(u_{+}-u_{-})}{(x_{+}-x_{-})^2}(\xi^{(+)}-\xi^{(-)})\left|_{E=\Omega=0}
\right.=0\, .
\label{eq:4-4}
\end{equation}
The coefficients in the prolongation satisfy

\begin{eqnarray}
\phi^{t}&=&\phi_{t}+(\phi_{u}-\tau_{t})u_{t}-\xi_{t}u_{x}-
\xi_{u}u_{t}u_{x}-\tau_{u}u_{t}^2
\label{eq:4-5} \\*[2ex]
\phi^{(\pm)}&=&\phi\left(x_{\pm},t,u(x_{\pm},t)\right).
\label{eq:4-6}
\end{eqnarray}
We substitute (\ref{eq:4-3}), (\ref{eq:4-5}) and (\ref{eq:4-6}) into
eq.(\ref{eq:4-4}) and eliminate $u_{t}(x,t)$ and $x_{+}$ using
the equations (\ref{eq:4-1a}) and (\ref{eq:4-1b}).
The only term involving $u_{x}$ is in
$\phi^{t}$. Its coefficient $\xi_{t}$ must vanish and we find $\dot{a}=\dot{b}=0$
in
the expression (\ref{eq:4-3}).

The remaining determining equation is

\begin{equation}
\begin{array}{l}
\left\{\phi_{t}+[\phi-u(\phi_{u}-\tau_{t}-au)]\frac{u_{+}-u_{-}}{x_{+}-x_{-}}
\right.\\*[2ex]
+\left.\frac{u}{x_{+}-x_{-}}\left[\phi(x_{+},t,u(x_{+},t))-
\phi(x_{-},t,u(x_{-},t))\right]\right\}_{x_{+}=2x-x_{-}}=0\, .
\end{array}
\label{eq:4-8}
\end{equation}
We differentiate twice with respect to $u_{+}$ and obtain $\phi_{u_{+}u_{+}}=0$,
so
that we have $\phi(x,t,u)=\phi_{1}(x,t)u+\phi_{0}(x,t)$. Substituing back into
eq.(\ref{eq:4-8}) we obtain the final result, namely

\begin{equation}
\xi=ax+b\ \ ,\ \ \tau=c_{1}t+c_{2}\ \ ,\ \ \phi=(a-c_{1})u\, .
\label{eq:4-8e}
\end{equation}
Thus, the difference scheme (\ref{eq:4-1a}), (\ref{eq:4-1b}) which is the usual
Volterra equation, is invariant under a $4$-dimensional group of Lie point
transformations. The symmetry algebra is spanned by

\begin{equation}
\hat P_{0}=\partial_{t}\ \ ,\ \ \hat P_{1}=\partial_{x}\ \ ,\ \ \hat
D_{0}=t\partial{t}-u\partial_{u}\ \ ,\ \ \hat D_{1}=x\partial{x}+u\partial_{u}
\label{eq:4-9}
\end{equation}
(two translations and two dilatations).

The continuous limit of the system (\ref{eq:4-1a}), (\ref{eq:4-1b}) is the Euler
equation in $1+1$ dimensions

\begin{equation}
u_{t}+uu_{x}=0\, .
\label{eq:4-10}
\end{equation}
Its symmetry group is infinite-dimensional and can be obtained by standard
techniques
\cite[\ldots,8]{3} (though we have not found it given explicitely in the
literature). Its
symmetry algebra is spanned by

\begin{equation}
\begin{array}{c}
\hat X(\xi)=\xi(z,u)\partial_{x}\ \ ,\ \
\hat T(\tau)=\tau(z,t,u)\left(\partial_{t}+u\partial_{x}\right) \\*[2ex]
\hat F(\phi)=\phi(z,u)\left(t\partial_{x}+\partial_{u}\right)\ \ ,\ \
z=x-ut
\end{array}
\label{eq:4-11}
\end{equation}
where $\xi,\, \tau$ and $\phi$ are arbitrary functions of their arguments.

The Volterra equation (\ref{eq:4-1a}) is certainly not a `symmetry preserving'
discretization of the Euler equation (\ref{eq:4-10}) on a uniform lattice. It only
preserves the four-dimensional subalgebra (\ref{eq:4-9}) of the
infinite-dimensional symmetry algebra (\ref{eq:4-11}). Let us mention
here that eq.(\ref{eq:4-1a}) is well known to be a bad numerical
scheme for eq.(\ref{eq:4-10}).

\subsection{A general nearest neighbour interaction equation}

Let us consider the difference scheme

\begin{eqnarray}
E&\equiv& u_{tt}-F(t,x_{+},x,x_{-},u_{+},u,u_{-})=0\, ,
\label{eq:4-12a} \\*[2ex]
\Omega&\equiv& x_{+}-2x+x_{-}=0\, ,
\label{eq:4-12b}
\end{eqnarray}
where $F$ is an arbitrary smooth function satisfying

\begin{equation}
(F_{u_{+}},F_{u_{-}})\neq (0,0)\, .
\label{eq:4-13}
\end{equation}
A symmetry analysis of a similar class of equations was recently performed for a
fixed
(non transformable) regular lattice \cite{12}. More specifically, the assumption
was
$x_{n}=n,\, n\in Z$.

The prolongation formula for the vector field (\ref{eq:2-22}) is
(\ref{eq:2-23}),...,(\ref{eq:2-26}). Applying it to eq.(\ref{eq:4-12b}) we obtain
that $\xi$ has the form (\ref{eq:4-3}), just as for the Volterra equation.
Applying
the
prolongation to the eq.(\ref{eq:4-12a}) we obtain

\begin{equation}
\phi^{tt}-\tau F_{t}-(ax+b) F_{x}-(ax_{+}+b) F_{x_{+}}-(ax_{-}+b) F_{x_{-}}
-\phi F_{u}-\phi^{(+)} F_{u_{+}}-
\phi^{(-)}
F_{u_{-}}\left|_{E=\Omega=0}\right.=0\, .
\label{eq:4-14}
\end{equation}
We substitute the expression for $\phi^{tt},\, \phi^{(+)}$ and $\phi^{(-)}$ and
set
the coefficients of $u_{t}^3$, $u_{t}^2$, $u_{t}^2u_{x}$,
$u_{t}u_{xt}$, $u_{xt}$, $u_{t}$  equal to zero, after eliminating
$u_{tt}$ and $x_{+}$,
using equations (\ref{eq:4-12a}), (\ref{eq:4-12b}). The result is that for any
interaction $F$ satisfying condition (\ref{eq:4-13}), we have

\begin{equation}
\tau=\tau(t)\ \ ,\ \ \xi=ax+b\ \ ,\ \
\phi=\left[\frac{\dot{\tau}}{2}+\alpha(x)\right]u+B(x,t)\, .
\label{eq:4-15}
\end{equation}
The as yet unspecified functions $\tau(t),\, \alpha(x),\, B(x,t)$ and constants
$a,\,
b$ satisfy a remaining determining equation, namely

\begin{equation}
\begin{array}{l}
\left\{\frac{1}{2}\tau_{ttt}u+B_{tt}-(\frac{3}{2}\tau_{t}-\alpha)F+\tau F_{t}
-(ax+b) F_{x}-(ax_{+}+b) F_{x_{+}} \right. \\*[2ex]
-(ax_{-}+b) F_{x_{-}}-\left[(\frac{1}{2}\tau_{t}+\alpha(x))u+B\right]
F_{u}-\left[(\frac{1}{2}\tau_{t}+\alpha(x_{+}))u_{+}+B(x_{+},t)\right] F_{u_{+}}
\\*[2ex]
\left.-\left[(\frac{1}{2}\tau_{t}+\alpha(x_{-}))u_{-}+B(x_{-},t)\right]
F_{u_{-}}\right\}_{x_{+}=2x-x_{-}}=0\, .
\end{array}
\label{eq:4-16}
\end{equation}

The results (\ref{eq:4-15}), (\ref{eq:4-16}) agree with those of Ref.\cite{12},
but
are
more general. The reason for the increase in generality is that here the lattice
is
not fixed {\em a priori} and hence the vector field (\ref{eq:2-22}) contains a
term
proportional to $\partial_{x}$.

To proceed further, we restrict the interaction $F$ to have a specific form.

\subsection{Equation with $F=(x_{+}-x)^{6}(u_{+}-2u+u_{-})^{-3}$}

Let us consider a special case of the system (\ref{eq:4-12a}), (\ref{eq:4-12b}),
namely

\begin{eqnarray}
u_{tt}=\frac{(x_{+}-x)^{6}}{(u_{+}-2u+u_{-})^{3}} ,
\label{eq:4-17a} \\*[2ex]
x_{+}-2x+x_{-}=0 .
\label{eq:4-17b}
\end{eqnarray}
We substitute $F$ of eq.(\ref{eq:4-17a}) into the determining equation
(\ref{eq:4-16}) and clear the denominator. The dependence on $u,\, u_{+}$ and
$u_{-}$
is explicit and we obtain

\begin{equation}
\begin{array}{l}
\tau_{ttt}=0\ \ ,\ \ B_{tt}=0\ \ ,\ \
B(x_{+},t)-2B(x,t)+B(x_{-},t)=0 , \\*[2ex]
\alpha(x)(x_{+}-x)+6(ax+b)-6(ax_{+}+b)+3\alpha(x_{+})(x_{+}-x)=0\, .
\end{array}
\label{eq:4-18}
\end{equation}
Analysing the system (\ref{eq:4-18}) in the usual manner, we obtain a
9-dimensional Lie
algebra with
basis

\begin{equation}
\begin{array}{c}
\hat P_{0}=\partial_{t}\ ,\ \hat P_{1}=\partial_{x}\ ,\
\hat D_{1}=2t\partial_{t}+u\partial{u}\ ,\
\hat D_{2}=2x\partial_{x}+3u\partial{u} , \\*[2ex]
\hat C=t^2\partial_{t}+tu\partial{u}\ ,\
\hat W_{1}=\partial{u}\ ,\
\hat W_{2}=t\partial{u}\ ,\
\hat W_{3}=x\partial{u}\ ,\
\hat W_{4}=tx\partial{u}\, .
\end{array}
\label{eq:4-19}
\end{equation}

A related system was studied earlier \cite{12}, namely

\begin{equation}
\ddot{u}_{n}(t)=\left[(\gamma_{n}-\gamma_{n-1})u_{n+1}+(\gamma_{n+1}-\gamma_{n-1})
u_{
n}+(\gamma_{n-1}-\gamma_{n})u_{n+1}\right]^{-3} ,
\label{eq:4-20}
\end{equation}
where $\gamma_{n}$ is any function of $n$, satisfying $\gamma_{n+1}\neq
\gamma_{n}$.
If we take $\gamma_{n}=n$ in eq.(\ref{eq:4-20}) and $x=n$ in (\ref{eq:4-17a}),
(\ref{eq:4-17b}) the two systems coincide. The symmetry algebra found in
Ref.\cite{12}
is the subalgebra $\{\hat P_{0}\, ,\, \hat D_{1}\, ,\, \hat C\, ,\, \hat W_{1}\,
,\,
\hat W_{2}\, ,\, \hat W_{3}\, ,\, \hat W_{4}\}$ of the algebra (\ref{eq:4-19}).
The
elements $\hat P_{1}$ and $\hat D_{2}$ are absent, since the lattice is fixed.
Shifts
$n'=n+N$ are allowed, but are not infinitesimal.

The system (\ref{eq:4-17a}), (\ref{eq:4-17b}) has a continuous limit

\begin{equation}
u_{tt}=\frac{1}{u_{xx}^3}\, .
\label{eq:4-20e}
\end{equation}
The symmetry algebra of eq.(\ref{eq:4-20e}) coincides with (\ref{eq:4-19}), i.e.
the
system (\ref{eq:4-17a}), (\ref{eq:4-17b}) is a symmetry preserving discretization
of
eq.(\ref{eq:4-20e}). We emphasize that eq.(\ref{eq:4-17a}) was obtained as part of
a
classification of difference equations \cite{12}, not in any connection with the
PDE
(\ref{eq:4-20e}).

\subsection{Equation without a continuous limit}

Let us now consider another special case of the system (\ref{eq:4-12a}),
(\ref{eq:4-12b}), namely

\begin{eqnarray}
u_{tt}=\frac{1}{(u_{+}-2u+u_{-})^{3}}  &,& x_{+}-2x+x_{-}=0\, .
\label{eq:4-21}
\end{eqnarray}

Substituing for $F$ into eq.(\ref{eq:4-16}) and proceeding as in Section~4.3. we
again obtain a 9-dimensional symmetry algebra. It differs from that given in
eq.(\ref{eq:4-19}) only in that $D_{2}$ is replaced by
$\tilde D_{2}=x\partial{x}$. For $h=x_{+}-x$ satisfying $h\rightarrow 0$ we find
$u_{tt}$ finite, but $(u_{+}-2u+u_{-})^{-3}\rightarrow \infty$, so the limit
$h\rightarrow 0$ does not exist.

\section{Conclusions}

The main questions to be addressed in a program aiming at using Lie group theory
to
solve difference equations are: (i) How does one define the symmetries? (ii) How
does
one calculate the symmetries? (iii) What does one do with the symmetries?

In this article we define the symmetries as in eq.(\ref{eq:2-11}), that is we
consider
only Lie point transformations that act simultaneously in a difference equation
(\ref{eq:E=0}) and lattice equation (\ref{eq:omega=0}). The fact that the lattice
also transforms is in the spirit of Dorodnitsyn's approach to discretizing
differential equations. In most symmetry studies of difference equations
\cite[\ldots, 26]{9}
the
lattice is fixed and nontransformable, e.g. given by the equation $x=n,\, n\in
Z$. For nontransforming lattices we need to go beyond point symmetries
to
catch transformations of interest\cite{17}.

Once the class of symmetries that we wish to consider is defined, the matter of
calculating them becomes purely technical. We proposed an algorithm for
calculating
symmetries in Section~2 (see eq. (\ref{eq:2-22}),...,(\ref{eq:2-27})) and applied
it
in Section~3 and 4. Symmetry algorithms for fixed lattices were presented
elsewhere \cite[-, 14]{10}.

Equations (\ref{eq:4-17a}) and (\ref{eq:4-20}) provide good examples of different
approaches. The symmetry algebra (\ref{eq:4-19}) of the system (\ref{eq:4-17a}),
(\ref{eq:4-17b}) happens to coincide with the symmetry algebra of the continuous
limit (\ref{eq:4-20e}). The symmetry algebra of the related equation
(\ref{eq:4-20})
was calculated elsewhere \cite{12}. It is a 7-dimensional subalgebra of
the algebra (\ref{eq:4-19}), obtained by dropping $\hat P_{1}$ and $\hat D_{2}$.
It was
obtained by the `intrinsic method' \cite{11}. The symmetry algebra of eq.(\ref{eq:4-20}) can
also be obtained from that of the system (\ref{eq:4-17a}), (\ref{eq:4-17b}) by
taking
a specific solution $x=n$ of eq.(\ref{eq:4-17b}) and reducing the algebra
(\ref{eq:4-19}) to the one that preserves this solution.

As far as applications of symmetries are concerned, they are the same for
differential equations and difference ones, in particular, symmetry reduction.

First, consider translationally invariant solutions, i.e. solutions invariant
under
the subgroup generated by $\hat X=\hat P_{0}-v \hat P_{1}$ with $v$ constant and
$\hat P_{0},\, \hat P_{1}$ as in eq.(\ref{eq:4-19}). We find that the solution,
the
differential-difference equations (D$\Delta$E) (\ref{eq:4-17a}), (\ref{eq:4-17b})
and
the PDE (\ref{eq:4-21}) reduce to

\begin{eqnarray}
u(x,t)=G(\eta)\ \ ,\ \ \eta=x+vt
\label{eq:5-1} \\*[2ex]
v^{2}G_{\eta\eta}[G(\eta+h)-2G(\eta)+G(\eta-h)]^{3}=h^{6}
\label{eq:5-2} \\*[2ex]
v^{2}G_{\eta\eta}^{4}=1
\label{eq:5-3}
\end{eqnarray}
respectively. Surprisingly, the difference equation (\ref{eq:5-2}) and the ODE
(\ref{eq:5-3}) have exactly the same solution for all values of the spacing $h$,
namely

\begin{equation}
G=\pm\frac{1}{2\sqrt{v}}\eta^2+A\eta+B\ \ ,\ \ v\neq 0
\label{eq:5-4}
\end{equation}
where $A$ and $B$ are integration constants. Thus, the system (\ref{eq:4-17a}),
(\ref{eq:4-17b}) is not only a symmetry preserving discretization. It also
preserves
translationally invariant solutions.

As a second example, consider solutions invariant under dilations generated by
$\hat D_{1}$ of eq.(\ref{eq:4-19}).The reduction formula, reduced D$\Delta$E and
reduced PDE are

\begin{eqnarray}
u(x,t)=t^{1/2}G(x)
\label{eq:5-5} \\*[2ex]
G(x)\, [G(x+h)-2G(x)+G(x-h)]^{3}=-4h^{6}
\label{eq:5-6} \\*[2ex]
G\, G_{xx}^{3}=-4
\label{eq:5-7}
\end{eqnarray}
respectively. A particular solution of eq.(\ref{eq:5-7}) is
$G(x)=4(-3)^{-3/4}(x-x_{0})^{3/2}$. This is not an exact solution of
eq.(\ref{eq:5-6}), but the solution of (\ref{eq:5-6}) and (\ref{eq:5-7}) coincide
to
order $h^{2}$, rather than just $h$.

As a final example of symmetry reduction, consider the subgroup corresponding to
$\hat D_{2}-3\hat D_{1}$ of eq.(\ref{eq:4-19}). The reduction formulas are

\begin{eqnarray}
u(x,t)=G(\eta)\ \ ,\ \ \eta=x^{3}t
\label{eq:5-8} \\*[2ex]
G_{\eta\eta}=\frac{(\eta_{+}^{1/3}-\eta^{1/3})^{6}}{\eta^{2}[G(\eta_{+})-2G(\eta)
+G(\eta_{-})]^{3}}\ \ ,\ \
\eta_{+}^{1/3}-2\eta^{1/3}+\eta_{-}^{1/3}=0 \label{eq:5-9}  \\*[2ex]
27\eta^{3}G_{\eta\eta}[3\eta G_{\eta\eta}+2G_{\eta}]^{3}=1
\label{eq:5-10}
\end{eqnarray}
While we are not able to solve the ODE (\ref{eq:5-10}), nor the difference scheme
(\ref{eq:5-9}), we see that in both cases we get a reduction of the number of
independent variables. We mention that this last reduction would not be obtained
on a fixed
lattice.

Let us sum up the situation with this particular approach to symmetries of
difference
equations.

\begin{enumerate}
\item Lie point symmetries acting simultaneously on given equations and lattices
can
be calculated using the reasonably simple algorithm presented in this article.

\item Symmetries can be used to perform symmetry reduction for D$\Delta$E.
\end{enumerate}

Work is in progress on other applications of symmetries of discrete equations, in
particular solving ordinary difference equations.

\section*{Acknowledgments}

The authors thank V. Dorodnitsyn, R. Kozlov and S. Lafortune for stimulating
discussions. The research of PW is partially supported by grants from NSERC of
Canada
and FCAR du Qu\'ebec. The research reported here is also partly supported by the
NATO
grant CRG960717 and a Cultural Agreement Universit\`a Roma Tre--Universit\'e de
Montr\'eal. ST and PW thank the Universit\`a Roma Tre for hospitality, DL similarly
thanks the Centre de Recherches Math\'ematiques, Universit\'e de Montr\'eal.

\newpage

\end{document}